\documentclass[aps,prd,preprint,numerical,showpacs,showkeys]{revtex4-1}

\usepackage{amsmath}
\usepackage{amssymb}
\usepackage{amsfonts}
\usepackage{graphicx}
\usepackage{dcolumn}
\usepackage{bm}
\usepackage{hyperref}
\hypersetup{colorlinks=false}

\begin{document}

\title[Wheeler-DeWitt equation in the Friedmann-Robertson-Walker universe]{Class of solutions of the Wheeler-DeWitt equation in the Friedmann-Robertson-Walker universe}

\date{\today}

\author{H. S. Vieira}
\email{horacio.santana.vieira@hotmail.com}
\affiliation{Departamento de F\'{i}sica, Universidade Federal da Para\'{i}ba, Caixa Postal 5008, CEP 58051-970, Jo\~{a}o Pessoa, PB, Brazil}
\affiliation{Centro de Ci\^{e}ncias, Tecnologia e Sa\'{u}de, Universidade Estadual da Para\'{i}ba, CEP 58233-000, Araruna, PB, Brazil}
\author{V. B. Bezerra}
\email{valdir@fisica.ufpb.br}
\affiliation{Departamento de F\'{i}sica, Universidade Federal da Para\'{i}ba, Caixa Postal 5008, CEP 58051-970, Jo\~{a}o Pessoa, PB, Brazil}

\begin{abstract}
We show that the solutions of the Wheeler-DeWitt equation in a homogeneous and isotropic universe are given by triconfluent Heun functions for the spatially closed, flat, and open geometries of the Friedmann-Robertson-Walker universe filled with different forms of energy. In a matter-dominated universe, we find the polynomial solution and the energy density spectrum. In the cases of radiation-dominated and vacuum universes, there are no polynomial solutions as shown.
\end{abstract}



%\preprint{AIP/123-QED}

\maketitle

%\begin{quotation}
%...
%\end{quotation}

%
%%%%%%%%%%%%%%%%%%%%%%%%%%%%%%%%%%%%%%%%%%%%%%%%%%%%%%%%%%%%%%%%%%%%%%%%%%%%%%%%%%%%%%%%%%%%%% Introduction
%
\section{Introduction}
The first approach based on the application of the quantum theory to describe the universe was presented in the later 1960s by Wheeler \cite{AnnPhys.2.604} and DeWitt \cite{PhysRev.160.1113}. At that time, they proposed a quantum gravity equation to describe the wave function of the universe and its evolution, which is known as the Wheeler-DeWitt (WDW) equation \cite{Wheeler:1968}. This equation is analogous to a zero-energy Schr\"{o}dinger equation of which the Hamiltonian could contain the gravitational field as well as nongravitational fields, as for example, scalar fields. If these fields are present in the Hamiltonian, the dynamical variables are the scale factor and the scalar field as well as their respective conjugate momenta.

In the case in which only the gravitational field is present, the solutions of the WDW equation defined in the minisuperspace depend just on a unique parameter, namely, the scale factor, and thus the properties of the wave functions will be characterized exclusively on this parameter. It is worth calling attention to the fact that the WDW equation leads us to stationary wave functions, due to the fact that it does not contain any classical parameter which can be identified with this quantity.

From that time up to now, this equation has inspired a lot of investigations in quantum cosmology \cite{PhysRevD.25.2065,PhysRevD.28.2960,PhysRevD.33.3560}, in which context the evolution of the universe is determined by the quantum states which obey the WDW equation \cite{EurPhysJC.57.769,PhysLettB.661.37,PhysRevD.86.063504,PhysRevD.88.084012,PhysLettB.748.361}, in such a way that in the appropriate classical limit the Friedmann solutions are recovered. Along this line of research, some results using different approaches have been obtained \cite{JHEP.04.026,GenRelativGravit.38.1645,PhysRevD.77.066017,PhysLettB.725.463,ClassQuantumGrav.31.095011,PhysRevD.90.124076,PhysRevD.92.126010,NuclPhysB.905.313}. Especially, we emphasize the line of research which uses the WDW equation as a fundamental tool to formulate the so-called loop quantum gravity \cite{PhysRevLett.57.2244}.

In spite of all problems present in the original formulation of the WDW equation, such as the breaking of relativistic covariance, the absence of a time variable, and others \cite{NuclPhysB.264.185}, nowadays, it is considered a fundamental tool to elaborate some models to explain the quantum properties of the gravitational phenomena, taking together quantum mechanics and general relativity \cite{ClassQuantumGrav.32.124005} and for this reason the work on quantum cosmology based on the WDW equation is going on.

The WDW equation does not provide the best way to describe the quantum gravity scenario, but it has opened up a route to describe a theory of quantum gravity and for this reason constitutes a powerful conceptual element to construct a theory in which the quantum dynamics of the gravitational phenomena can be explained consistently \cite{ClassQuantumGrav.32.124005}. Thus, it is certainly important to find solutions of this equation and use these to learn about the possibility to construct a consistent theory that takes into account the quantum features of our Universe.

This paper is organized as follows. In Sec. II, the WDW equation is presented. In Sec. III, we obtain its solution for a universe filled with matter. Sections IV and V present the solution for the three possible geometries and a universe filled out with radiation and vacuum energy, respectively. In Sec. VI, the conclusions are presented.
%
%%%%%%%%%%%%%%%%%%%%%%%%%%%%%%%%%%%%%%%%%%%%%%%%%%%%%%%%%%%%%%%%%%%%%%%%%%%%%%%%%%%%%%%%%%%%%% WDW equation in Friedmann-Robertson-Walker universe
%
\section{WDW equation in Friedmann-Robertson-Walker universe}
The WDW equation was derived originally in the minisuperspace approximation for closed and empty universes \cite{PhysRevD.25.2065,PhysRevD.28.2960}. But it is important to consider different geometries as well as the possibility to have a presence of matter and radiation \cite{EurJPhys.19.143} and others forms of energy.

Let us assume that the universe is homogeneous and isotropic and therefore can be described using the minisuperspace model \cite{ClassQuantumGrav.30.143001} with one fundamental parameter, the scale factor $a$ ($0 \leq a < \infty$) of the Friedmann-Robertson-Walker (FRW) universe.

In the case of the FRW universe, the Lagrangian takes the form \cite{IntJModPhysD.11.527}
\begin{equation}
L=-\frac{3 \pi c^{2}}{4G}a^{3}\left[\left(\frac{\dot{a}}{a}\right)^{2}-\frac{k c^{2}}{a^{2}}+\frac{8 \pi G}{3 c^{2}}(\rho+\rho_{vac})\right]\ ,
\label{eq:Lagrangian_WDW}
\end{equation}
where different forms of energy are taken into account, as well as geometries ($k=-1,0,+1$, which corresponds to the open, flat, and closed universes, respectively).

Substituting the momentum conjugate to $a$, given by
\begin{equation}
p=\frac{\partial L}{\partial \dot{a}}=-\frac{3 \pi c^{2}}{2G}a\dot{a}\ ,
\label{eq:momentum_WDW}
\end{equation}
and the Lagrangian into the Euler-Lagrange equation $\dot{p}-\partial L/\partial a=0$, we get the following result:
\begin{equation}
\left(\frac{\dot{a}}{a}\right)^{2}-\frac{k c^{2}}{a^{2}}-\frac{8 \pi G}{3 c^{2}}(\rho+\rho_{vac})=0\ .
\label{eq:result_WDW}
\end{equation}
The classical evolution of the universe at different stages can be obtained by solving this equation, which can be used to conclude that the Hamiltonian $H=p\dot{a}-L$, written in terms of the momentum, $p$,
\begin{equation}
H(p,a)=\frac{3 \pi c^{2}}{4G}a^{3}\left[\frac{4G^{2}p^{2}}{9 \pi^{2} c^{4} a^{4}}+\frac{k c^{2}}{a^{2}}-\frac{8 \pi G}{3 c^{2}}(\rho+\rho_{vac})\right] \equiv 0\ ,
\label{eq:Hamiltonian_WDW}
\end{equation}
is identically zero.

Now, in order to obtain the WDW equation in the minisuperspace approximation, we make the replacement $p \rightarrow -i\hbar\partial/\partial a$ and impose that $H\Psi=0$. Thus, we get the WDW equation in a suitable minisuperspace by substituting the dynamical variable and its conjugate by corresponding operators. Explicitly, in the FRW universe, the WDW equation reads \cite{IntJModPhysD.11.527}
\begin{equation}
\left\{\frac{d^{2}}{da^{2}}-\frac{9\pi c^{4}a^{2}}{4\hbar^{2}G^{2}}\left[kc^{2}-\frac{8\pi Ga^{2}}{3c^{2}}(\rho+\rho_{vac})\right]\right\}\Psi(a)=0\ ,
\label{eq:WDE}
\end{equation}
where we are considering that the wave function depends only on the scale factor.

Let us consider the expression for the energy density
\begin{equation}
\rho=A_{\omega}a^{-3(\omega+1)}\ ,
\label{eq:WDE_density}
\end{equation}
where $A_{\omega}=\rho_{\omega 0}a_{0}^{3(\omega+1)}$ with $\rho_{\omega 0}$ being the value of $\rho_{\omega}$ at present time, and $\omega$ is such that
\begin{equation}
\omega=\left\{
\begin{array}{rl}
	0 & \mbox{for dust (matter predominance}; \rho_{m})\ ,\\
	\frac{1}{3} & \mbox{for radiation (radiation predominance}; \rho_{r})\ ,\\
	-1 & \mbox{for de Sitter (false vacuum}; \rho_{v})\ .
\end{array}
\right.
\label{eq:WDE_omega}
\end{equation}
The energy density of the vacuum, $\rho_{vac}$, can be expressed, in terms of the cosmological constant, in the following form:
\begin{equation}
\rho_{vac}=\frac{\Lambda c^{4}}{8\pi G}\ .
\label{eq:WDE_energy_density_vacuum}
\end{equation}

Now, substituting Eqs.~(\ref{eq:WDE_density})--(\ref{eq:WDE_energy_density_vacuum}) into Eq.~(\ref{eq:WDE}), we obtain
\begin{equation}
-\hbar^{2}\frac{d^{2}\Psi(a)}{da^{2}}+V_{eff}(a)\Psi(a)=0\ ,
\label{eq:WDE_mov_1}
\end{equation}
where
\begin{equation}
V_{eff}(a)=\frac{9\pi^{2}c^{6}k}{4G^{2}}\left[a^{2}-\frac{1}{k}\left(\frac{\Lambda}{3}a^{4}+\frac{8 \pi G A_{\omega}}{3 c^{4}}a^{1-3\omega}\right)\right]\ .
\label{eq:WDE_effective_potential_energy_WDE}
\end{equation}
Equation (\ref{eq:WDE_mov_1}) looks like a one-dimensional time-independent Schr\"{o}dinger equation for energy zero and for a particle with $1/2$ of the unit mass, with an effective potential $V_{eff}(a)$ given by Eq.~(\ref{eq:WDE_effective_potential_energy_WDE}), which is valid for arbitrary $\omega$.

The expression for the effective potential, $V_{eff}$, tells us that the minisuperspace can be divided into regions where $V_{eff} > 0$ and $V_{eff} < 0$, the limits of which depend on the signal of the cosmological constant, $\Lambda$.

Note that for small values of $a$ we can neglect the terms proportional to $a^{4}$, and thus, in a de Sitter scenario, the effective potential is proportional to $a^{2}$, in this limit. In this case, the solution of the WDW equation is given in terms of Bessel functions \cite{ClassQuantumGrav.30.143001}, and the wave function approaches a constant for $a \rightarrow 0$. On the other hand, for arbitrary values of $a$, the potential is not so simple, and as a consequence, the solutions of the WDW equation are more complicated and given in terms of Heun functions \cite{Ronveaux:1995} as we will show in the next sections, in which we will obtain the analytical solutions of Eq.~(\ref{eq:WDE_mov_1}) with an effective potential given by Eq.~(\ref{eq:WDE_effective_potential_energy_WDE}), for any value of the scale factor.

Note that when $a=0$ the universe corresponds to a quantum FRW universe with zero radius. In this case, the solution of the WDW equation approaches a constant and describes a state called ``nothing'' in the literature \cite{PhysLettB.117.25}. This state can be created by quantum mechanically tunnelling through the potential barrier that appears at $a=a_{0} \neq 0$.
%
%%%%%%%%%%%%%%%%%%%%%%%%%%%%%%%%%%%%%%%%%%%%%%%%%%%%%%%%%%%%%%%%%%%%%%%%%%%%%%%%%%%%%%%%%%%%%% Solution of the WDW equation and energy density spectrum for \texorpdfstring{$\omega=0$}{w=0}
%
\section{Solution of the WDW equation and energy density spectrum for \texorpdfstring{$\omega=0$}{w=0}}
In what follows, we will find the solutions of the WDW equation for arbitrary scale factor ($0 \leq a < \infty$), in terms of the Heun functions.

\begin{figure}%[htbp]
	%\centering
		\includegraphics[scale=0.50]{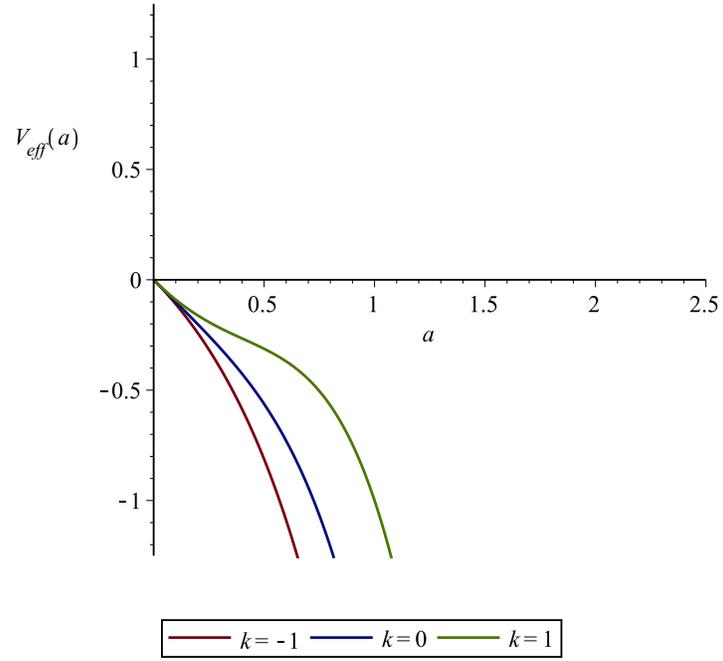}
	\caption{The effective potential energy, $V_{eff}(a)$, for $\omega=0$ and $\Lambda > 0$.}
	\label{fig:WDE_Fig1}
\end{figure}

\begin{figure}%[htbp]
	%\centering
		\includegraphics[scale=0.50]{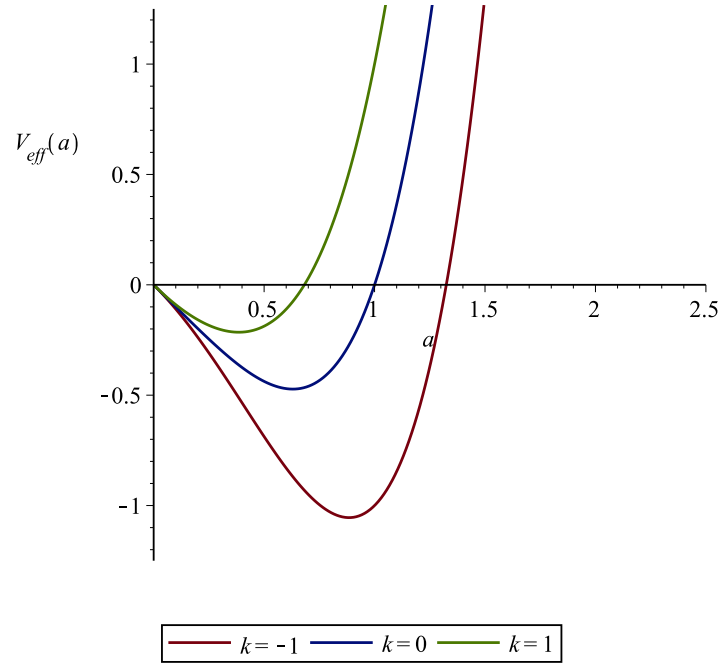}
	\caption{The effective potential energy, $V_{eff}(a)$, for $\omega=0$ and $\Lambda = -|\Lambda|$.}
	\label{fig:WDE_Fig2}
\end{figure}

First, we consider the solution of the WDW equation in the matter era, in which case $\omega=0$. Thus, Eq.~(\ref{eq:WDE_mov_1}) takes the form
\begin{equation}
-\hbar^{2}\frac{d^{2}\Psi(a)}{da^{2}}+\left(\frac{9\pi^{2}c^{6}k}{4G^{2}}a^{2}-\frac{6\pi^{3}c^{2}\rho_{m0}a_{0}^{3}}{G}a-\frac{3\pi^{2}c^{6}\Lambda}{4G^{2}}a^{4}\right)\Psi(a)=0\ .
\label{eq:WDE_mov_2}
\end{equation}
The behavior of $V_{eff}(a)$ for this case is shown in Figs.~\ref{fig:WDE_Fig1} and \ref{fig:WDE_Fig2}, for positive and negative values of the cosmological constant.

Note that for $\Lambda > 0$ the potential is unbounded (see Fig.~\ref{fig:WDE_Fig1}), differently from the empty universe case, where $V_{eff}(a)$ contains two terms, one proportional to $a^{2}$ and the other proportional to $a^{4}$, in such a way that there is a possibility to have bounded states as well as quantum states created by a tunnelling process \cite{EurJPhys.19.143}. For $\Lambda < 0$, we can have bounded states (see Fig.~\ref{fig:WDE_Fig2}).

It is more suitable to rewrite Eq.~(\ref{eq:WDE_mov_2}) in terms of dimensionless quantities. To do this, we first introduce the dimensionless parameter
\begin{equation}
\beta=\frac{6\pi^{2}\rho_{m0}a_{0}^{3}}{\hbar c \Omega}\ ,
\label{eq:WDE_beta_mov_2}
\end{equation}
where
\begin{equation}
\Omega=\left(-\frac{\Lambda}{3}\right)^{\frac{1}{2}}\ .
\label{eq:WDE_Omega_mov_2}
\end{equation}
Let us also use the dimensionless variable
\begin{equation}
x=\xi a\ ,
\label{eq:WDE_x}
\end{equation}
where the coefficient $\xi$ is given by
\begin{equation}
\xi=c\left(\frac{\pi\Omega}{\hbar G}\right)^{\frac{1}{3}}\ .
\label{eq:WDE_tau_mov_2}
\end{equation}
Then, Eq.~(\ref{eq:WDE_mov_2}) turns into
\begin{equation}
\frac{d^{2}\Psi(x)}{dx^{2}}+\left(\beta x-\frac{3}{2}\gamma x^{2}-\frac{9}{4}x^{4}\right)\Psi(x)=0\ ,
\label{eq:WDE_mov_3}
\end{equation}
where the parameter $\gamma$ is given by
\begin{equation}
\gamma=\frac{3c^{2}k}{2}\left(\frac{\pi}{\hbar G\Omega^{2}}\right)^{\frac{2}{3}}\ .
\label{eq:WDE_gamma_mov_3}
\end{equation}

Now, we can rewrite Eq.~(\ref{eq:WDE_mov_3}) in the form which resembles a Heun equation, by assuming that $\Psi$ can be written as \cite{Ronveaux:1995}
\begin{equation}
\Psi(x)=\mbox{e}^{-\frac{1}{2}(x^{3}+\gamma x)}P(x)\ ,
\label{eq:WDE_form_sol_mov_3}
\end{equation}
where $P(x)$ has to be determined appropriately. Substituting Eq.~(\ref{eq:WDE_form_sol_mov_3}) into Eq.~(\ref{eq:WDE_mov_3}), we conclude that $P(x)$ must satisfy the equation
\begin{equation}
\frac{d^{2}P}{dx^{2}}-(\gamma+3x^{2})\frac{dP}{dx}+\left[\frac{\gamma^{2}}{4}+(\beta-3)x\right]P=0\ ,
\label{eq:WDE_mov_4}
\end{equation}
the analytical solutions of which will be obtained in what follows.

Equation (\ref{eq:WDE_mov_4}) is a particular case of the triconfluent Heun equation (THE), which in the canonical form (referred to as the THE$_{1}$ equation) is given by \cite{AnnSocSciBruxelles.92.53}
\begin{equation}
\frac{d^{2}y}{dx^{2}}-(\gamma+3x^{2})\frac{dy}{dx}+[\alpha+(\beta-3)x]y=0\ ,
\label{eq:WDE_Triconfluent_Heun_Canonical}
\end{equation}
where $y(x)=\mbox{HeunT}(\alpha,\beta,\gamma;x)$ are the triconfluent Heun functions.

Let us assume that the solutions of Eq.~(\ref{eq:WDE_Triconfluent_Heun_Canonical}) can be written as \cite{JMathPhys.56.092501}
\begin{equation}
y(x)=\sum_{s=0}^{\infty}u_{s}x^{s}\ .
\label{eq:WDE_y(x)_Triconfluent_Heun}
\end{equation}
Thus, substituting Eq.~(\ref{eq:WDE_y(x)_Triconfluent_Heun}) into Eq.~(\ref{eq:WDE_Triconfluent_Heun_Canonical}), we get
\begin{eqnarray}
&& \sum_{s=2}^{\infty}s(s-1)u_{s}x^{s-2}-\sum_{s=1}^{\infty}\gamma su_{s}x^{s-1}-\sum_{s=1}^{\infty}3su_{s}x^{s+1}+\sum_{s=0}^{\infty}\alpha u_{s}x^{s}\nonumber\\
&& +\sum_{s=0}^{\infty}(\beta-3)u_{s}x^{s+1}=0\ ,
\label{eq:WDE_Triconfluent_Heun_Canonical_expansion_2}
\end{eqnarray}
from which we obtain the following recurrence relation of order 3:
\begin{eqnarray}
&& \sum_{S=-1}^{\infty}(S+3)(S+3)u_{S+3}x^{S+1}-\sum_{S=-1}^{\infty}\gamma(S+2)u_{S+2}x^{S+1}-\sum_{S=1}^{\infty}3Su_{S}x^{S+1}\nonumber\\
&& +\sum_{S=-1}^{\infty}\alpha u_{S+1}x^{S+1}+\sum_{S=0}^{\infty}(\beta-3)u_{S}x^{S+1}=0\ .
\label{eq:WDE_Triconfluent_Heun_Canonical_expansion_3}
\end{eqnarray}
Collecting all terms of the same order in $x$, we get
\begin{eqnarray}
&& [\alpha u_{0}-\gamma u_{1}+2u_{2}]\nonumber\\
&& +[(\beta-3)u_{0}+\alpha u_{1}-2\gamma u_{2}+6u_{3}]x\nonumber\\
&& +\sum_{S=1}^{\infty}[(\beta-3-3S)u_{S}+\alpha u_{S+1}-\gamma(S+2)u_{S+2}+(S+3)(S+2)u_{S+3}]x^{S+1}=0\ .
\label{eq:WDE_Triconfluent_Heun_Canonical_expansion_4}
\end{eqnarray}

Thus, Eq.~(\ref{eq:WDE_Triconfluent_Heun_Canonical_expansion_4}) gives the following recursion relations for the expansion coefficients:
\begin{eqnarray}
2u_{2}=\gamma u_{1}-\alpha u_{0}\ ,\nonumber\\
6u_{3}=2\gamma u_{2}+\alpha u_{1}-(\beta-3)u_{0}\ ,\nonumber\\
(S+3)(S+2)u_{S+3}=\gamma(S+2)u_{S+2}+\alpha u_{S+1}-(\beta-3-3S)u_{S}, \quad S \geq 0\ .
\label{eq:WDE_recursion_Triconfluent_Heun_expansion_4}
\end{eqnarray}
Considering appropriate choices of $u_{0}$ and $u_{1}$, and taking into account a convenient change in notation, we have
\begin{eqnarray}
u_{0}=1\ ,\nonumber\\
u_{1}=0\ ,\nonumber\\
u_{2}=-\frac{\alpha}{2}\ ,\nonumber\\
u_{3}=-\frac{1}{6}[\alpha\gamma+(\beta-3)]\ ,\nonumber\\
u_{s}=\frac{\gamma(s-1)u_{s-1}+\alpha u_{s-2}-(\beta+6-3s)u_{s-3}}{s(s-1)}, \quad s \geq 3\ ,
\label{eq:WDE_recursion_u_Triconfluent_Heun_expansion_4}
\end{eqnarray}
where $u_{-1}=0$.

Assuming that $(\alpha,\beta,\gamma) \in \mathbb{C}^{3}$, the triconfluent Heun functions can be written as
\begin{equation}
\mbox{HeunT}(\alpha,\beta,\gamma;x)=\sum_{s \geq 0}u_{s}(\alpha,\beta,\gamma)x^{s}\ ,
\label{eq:WDE_Triconfluent_Heun_expansion}
\end{equation}
where $u_{s}(\alpha,\beta,\gamma)$, given by Eq.~(\ref{eq:WDE_recursion_u_Triconfluent_Heun_expansion_4}), is a polynomial in the three variables $\alpha,\beta,\gamma$.

From the recursion relation given by Eq.~(\ref{eq:WDE_recursion_Triconfluent_Heun_expansion_4}), the function $\mbox{HeunT}(\alpha,\beta,\gamma;x)$ becomes a polynomial of degree $n$ if and only if the two following conditions are fulfilled \cite{Ronveaux:1995}:
\begin{equation}
\begin{array}{rl}
	\mbox{(i)}  & \beta=3(n+1), \quad n=0,1,2,\ldots\ ,\\
	\mbox{(ii)} & \Pi_{n+1}(\alpha,\gamma)=0\ ,
\end{array}
\label{eq:WDE_condiction_poly_Triconfluent_Heun}
\end{equation}
where $\Pi_{n+1}$ is a polynomial in $\alpha,\gamma$, with the degree in $\alpha$ being $n+1$.

The polynomial $\Pi_{n+1}(\alpha,\gamma)$ is the determinant of dimension $n+1$ given by
\begin{equation}
\left|
\begin{array}{ccccccccc}
	\alpha & -\gamma & 2 \cdot 1 & 0         & \ldots &           &           &              & 0        \\
	3n     & \alpha  & -2\gamma  & 3 \cdot 2 & \ldots &           &           &              &          \\
	0      & 3(n-1)  & \alpha    & -3\gamma  & \ldots &           &           &              &          \\
	0      & 0       & 3(n-2)    & \alpha    & \ldots &           &           &              &          \\
	\vdots & \vdots  & \vdots    & \vdots    & \ddots &           &           &              & \vdots   \\
	       &         &           &           &        & 3 \cdot 3 & \alpha    & -(n-1)\gamma & n(n-1)   \\
	       &         &           &           &        & 0         & 3 \cdot 2 & \alpha       & -n\gamma \\
	0      &         &           &           & \ldots & 0         & 0         & 3 \cdot 1    & \alpha   \\
\end{array}
\right|\ .
\label{eq:WDE_determinant_Triconfluent_Heun}
\end{equation}

Now, let us return to the wave functions $\Psi(x)$. Using Eq.~(\ref{eq:WDE_form_sol_mov_3}), we see that the physically acceptable solutions of Eq.~(\ref{eq:WDE_mov_4}) are given by
\begin{equation}
\Psi_{n}(x)=\mbox{e}^{-\frac{1}{2}(x^{3}+\gamma x)}P_{n}(\alpha,\gamma;x)\ ,
\label{eq:WDE_form_polyn_sol_mov_3}
\end{equation}
where the functions $P_{n}(\alpha,\gamma;x)$ are polynomials of degree $n$ satisfying Eq.~(\ref{eq:WDE_Triconfluent_Heun_Canonical}), with $\beta=3(n+1)$, $\Pi_{n+1}=0$, and $0 \leq s \leq n$, namely,
\begin{equation}
\frac{d^{2}P_{n}}{dx^{2}}-(\gamma+3x^{2})\frac{dP_{n}}{dx}+(\alpha+3nx)P_{n}=0\ .
\label{eq:WDE_Triconfluent_Heun_polynomials}
\end{equation}

The polynomials $P_{n}(\alpha,\gamma;x)=\mbox{HeunT}(\alpha,3(n+1),\gamma;x)$ are called Heun polynomials of the triconfluent case and are given by
\begin{equation}
P_{n}(\alpha,\gamma;x)=\sum_{s=0}^{n}u_{s}(\alpha,3(n+1),\gamma)x^{s}\ .
\label{eq:WDE_Triconfluent_Heun_polynomials_expansion}
\end{equation}
We believe that will be clear from the foregoing discussion that these polynomials are uniquely defined, except for an arbitrary multiplicative constant.

Now, let us substitute Eq.~(\ref{eq:WDE_Triconfluent_Heun_polynomials_expansion}) into Eq.~(\ref{eq:WDE_Triconfluent_Heun_polynomials}) in order to obtain the recurrence relation of order 2,
\begin{equation}
3u_{n-1}+\alpha u_{n}-\gamma(n+1)u_{n+1}+(n+1)(n+2)u_{n+2}=0\ ,
\label{eq:WDE_recursion_u_Triconfluent_Heun_polynomials_expansion}
\end{equation}
where $u_{-1}=0$. Note that the relation (\ref{eq:WDE_recursion_u_Triconfluent_Heun_polynomials_expansion}) with $n=0$ gives $u_{-1}=0$ if and only if $\Pi_{n+1}(\alpha,\gamma)=0$. If this is the case, computing recursively the $u_{s}$, in decreasing order, starting with $u_{n}=1$, we get
\begin{eqnarray}
u_{n-1}=-\frac{\alpha}{2}=\frac{(-1)^{1}}{1!3^{1}}\Pi_{1}(\alpha,\gamma)\ ,\nonumber\\
u_{n-2}=\frac{1}{18}(\alpha^{2}+3n\gamma)=\frac{(-1)^{2}}{2!3^{2}}\Pi_{2}(\alpha,\gamma)\ ,\nonumber\\
u_{n-3}=-\frac{1}{162}[\alpha^{3}+3n\alpha\gamma+6(n-1)\alpha\gamma+18n(n-1)]=\frac{(-1)^{3}}{3!3^{3}}\Pi_{3}(\alpha,\gamma)\ ,\nonumber\\
\vdots\nonumber\\
u_{n-s}=\frac{(-1)^{s}}{s!3^{s}}\Pi_{s}(\alpha,\gamma), \quad 0 \leq s \leq n\ ,
\label{eq:WDE_recursion_u_Triconfluent_Heun_polynomials_expansion_2}
\end{eqnarray}
where $\Pi_{0}(\alpha,\gamma)=1$. 

Now, we can rewrite Eq.~(\ref{eq:WDE_Triconfluent_Heun_polynomials_expansion}) as
\begin{equation}
P_{n}(\alpha,\gamma;x)=\sum_{s=0}^{n}\frac{(-1)^{s}}{s!3^{s}}\Pi_{s}(\alpha,\gamma)x^{n-s}\ .
\label{eq:WDE_Triconfluent_Heun_polynomials_expansion_2}
\end{equation}
The highest-order terms in this polynomial are given by
\begin{eqnarray}
&& x^{n}-\frac{1}{1!3^{1}}\alpha x^{n-1}+\frac{1}{2!3^{2}}(\alpha^{2}+3n\gamma)x^{n-2}-\frac{1}{3!3^{3}}[\alpha^{3}+3(3n-2)\alpha\gamma+18n(n-1)]x^{n-3}\nonumber\\
&& +Q(x)\ ,
\label{eq:WDE_polynomial_Triconfluent_Heun}
\end{eqnarray}
where $Q(x)$ is a polynomial with degree $\leq n-4$.

In our case, the parameter $\alpha$ takes the value $\gamma^{2}/4$,  and the parameter $\gamma$ is given by Eq.~(\ref{eq:WDE_gamma_mov_3}). Therefore, the expressions of the Heun polynomials $P_{n}(\alpha,\gamma;x)$, for $n=0,1,2$, are as follows:
\begin{itemize}
	\item $n=0$, $\beta=3$, $\Pi_{0+1}=0\ \Rightarrow\ \alpha=0$,
\end{itemize}
\begin{equation}
P_{0}(\alpha,\gamma;x)=1\ ,
\label{eq:WDE_P0}
\end{equation}

\begin{itemize}
	\item $n=1$, $\beta=6$, $\Pi_{1+1}=0\ \Rightarrow\ \alpha^{2}+3\gamma=0$,
\end{itemize}
\begin{equation}
P_{1}(\alpha,\gamma;x)=x-\frac{\alpha}{3}\ ,
\label{eq:WDE_P1}
\end{equation}

\begin{itemize}
	\item $n=2$, $\beta=9$, $\Pi_{2+1}=0\ \Rightarrow\ \alpha^{3}+12\alpha\gamma+36=0$,
\end{itemize}
\begin{equation}
P_{2}(\alpha,\gamma;x)=x^{2}-\frac{\alpha}{3}x+\frac{\alpha^{2}}{36}-\frac{1}{\alpha} \quad (\alpha \in \mathbb{C}^{*})\ .
\label{eq:WDE_P2}
\end{equation}

It is worth noting that for each value of $n$, namely, $n=0,1,2,\ldots$, corresponds a function $\mbox{HeunT}(\alpha,\beta,\gamma;x)$, which is a polynomial of degree $n$ in $x$.

To obtain the energy density spectrum corresponding to a solution of the WDW equation in a suitable minisuperspace, with $\omega=0$, let us use Eqs.~(\ref{eq:WDE_beta_mov_2}) and (\ref{eq:WDE_condiction_poly_Triconfluent_Heun}). Thus, we find the following result:
\begin{equation}
\rho_{m0,n}=(n+1)\frac{\hbar c \Omega}{2 \pi^{2} a_{0}^{3}}=(n+1)\frac{\hbar c}{2 \pi^{2} a_{0}^{3}}\left(-\frac{\Lambda}{3}\right)^{\frac{1}{2}}, \quad n=0,1,2,\ldots\ .
\label{eq:WDE_Energy_levels_WheelerDeWitt_general}
\end{equation}
If we consider a scenario in which the cosmological constant is negative, that is, $\Lambda=-|\Lambda|$, the energy spectrum can be rewritten as
\begin{equation}
\rho_{m0,n}=(n+1)\frac{\hbar c}{2 \pi^{2} a_{0}^{3}}\left(\frac{|\Lambda|}{3}\right)^{\frac{1}{2}}, \quad n=0,1,2,\ldots\ .
\label{eq:WDE_Energy_levels_Schrodinger}
\end{equation}

\begin{figure}%[htbp]
	%\centering
		\includegraphics[scale=0.50]{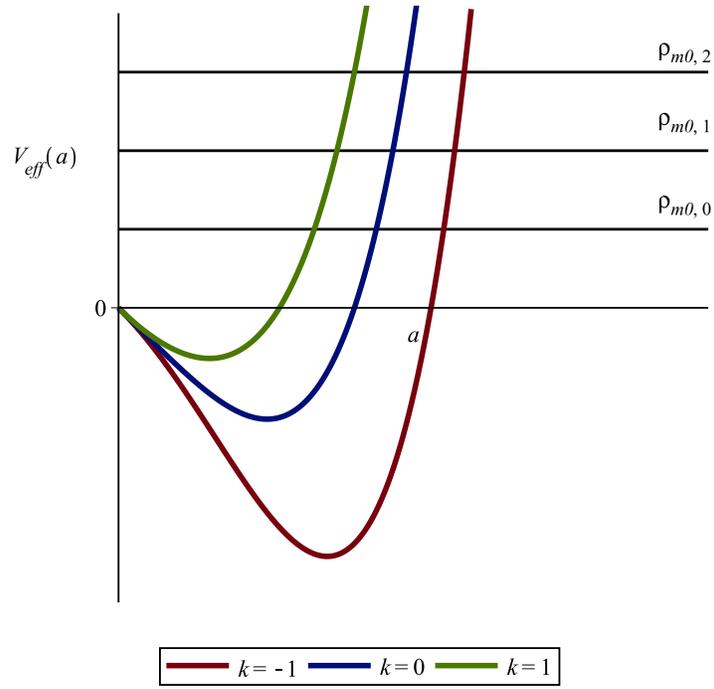}
	\caption{The effective potential energy, $V_{eff}(a)$, for $\omega=0$ and $\Lambda = -|\Lambda|$. The first three energy density levels are shown, too.}
	\label{fig:WDE_Fig3}
\end{figure}

We see from Eq.~(\ref{eq:WDE_Energy_levels_Schrodinger}) that this quantum mechanical energy spectrum consists of an infinite sequence of discrete levels (see Fig.~\ref{fig:WDE_Fig3}), which are equally spaced. Note that the eigenvalues given by Eq.~(\ref{eq:WDE_Energy_levels_Schrodinger}) are nondegenerate.

\begin{figure}%[htbp]
	%\centering
		\includegraphics[scale=0.50]{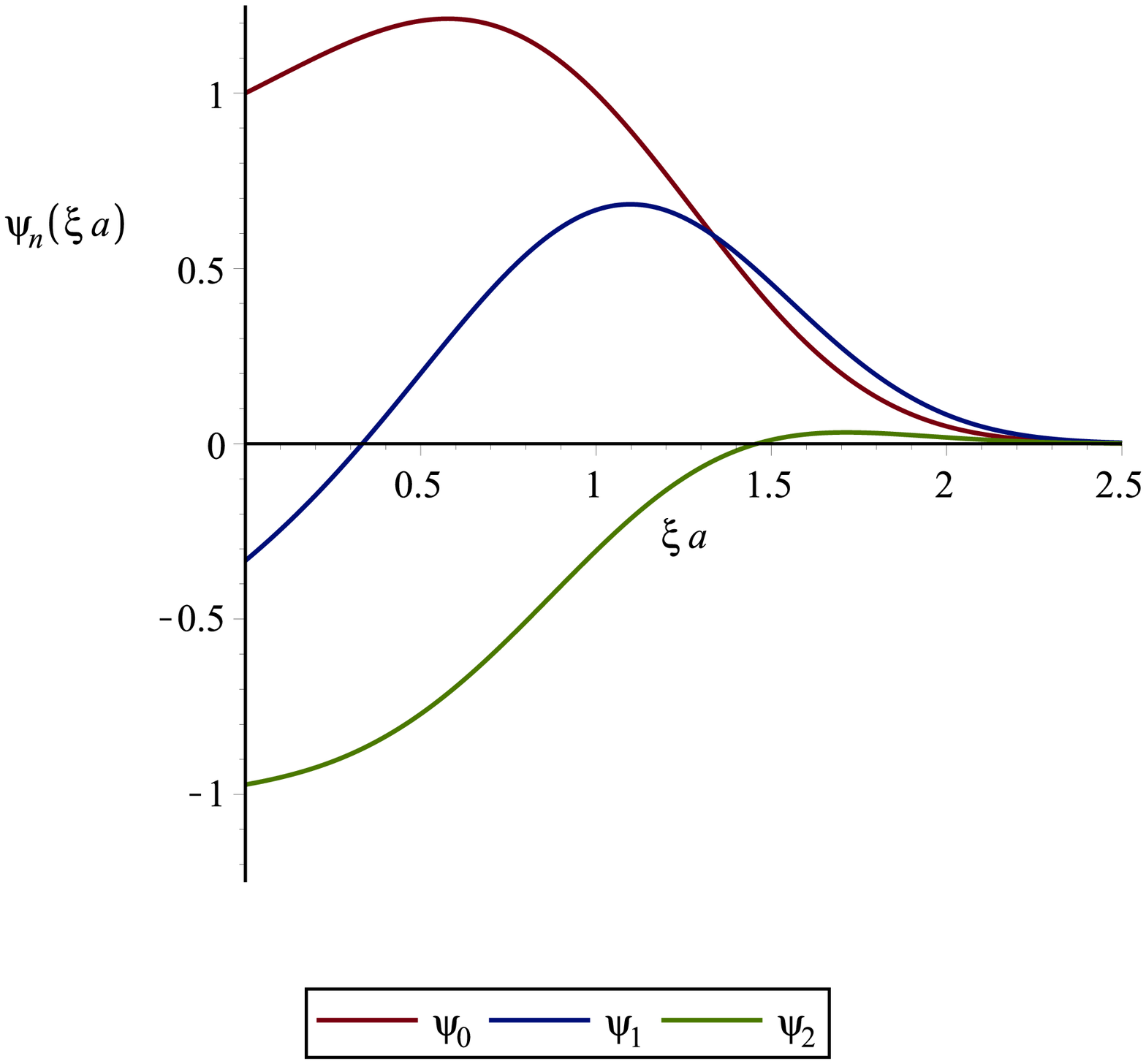}
	\caption{The first three wave functions for $k=-1$, $\omega=0$, and $\Lambda = -|\Lambda|$.}
	\label{fig:WDE_Fig4}
\end{figure}

Let us now turn to the original wave function. Using Eq.~(\ref{eq:WDE_form_polyn_sol_mov_3}), we see that to each discrete value $\rho_{m0,n}$, expressed by Eq.~(\ref{eq:WDE_Energy_levels_WheelerDeWitt_general}), there corresponds $n$ physically acceptable solutions (see Fig.~\ref{fig:WDE_Fig4}), given by
\begin{eqnarray}
\Psi_{n}(\xi a) & = & N_{n}\ \mbox{e}^{-\frac{1}{2}(\xi^{3} a^{3}+\gamma \xi a)}\ P_{n}(\alpha,\gamma;\xi a)\nonumber\\
& = & N_{n}\ \mbox{e}^{-\frac{1}{2}(\xi^{3} a^{3}+\gamma \xi a)}\ \mbox{HeunT}(\alpha,3(n+1),\gamma;\xi a)\ ,
\label{eq:WDE_form_polyn_sol_mov_4}
\end{eqnarray}
where the quantity $N_{n}$ is a constant to be determined and the parameters $\alpha$, $\beta$, and $\gamma$ are given by the following expressions:
\begin{equation}
\alpha=\frac{9c^{4}k^{2}}{16}\left(\frac{3\pi}{\hbar G \Lambda}\right)^{\frac{4}{3}}=\frac{9c^{4}k^{2}}{16}\left(\frac{3\pi}{\hbar G |\Lambda|}\right)^{\frac{4}{3}}\ ,
\label{eq:WDE_alpha_WheelerDeWitt}
\end{equation}
\begin{equation}
\beta=\frac{6\pi^{2}\rho_{m0}a_{0}^{3}}{\hbar c}\left(-\frac{3}{\Lambda}\right)^{\frac{1}{2}}=\frac{6\pi^{2}\rho_{m0}a_{0}^{3}}{\hbar c}\left(\frac{3}{|\Lambda|}\right)^{\frac{1}{2}}\ ,
\label{eq:WDE_beta_WheelerDeWitt}
\end{equation}
\begin{equation}
\gamma=\frac{3c^{2}k}{2}\left(-\frac{3\pi}{\hbar G \Lambda}\right)^{\frac{2}{3}}=\frac{3c^{2}k}{2}\left(\frac{3\pi}{\hbar G |\Lambda|}\right)^{\frac{2}{3}}\ .
\label{eq:WDE_gamma_WheelerDeWitt}
\end{equation}

\begin{figure}%[htbp]
	%\centering
		\includegraphics[scale=0.50]{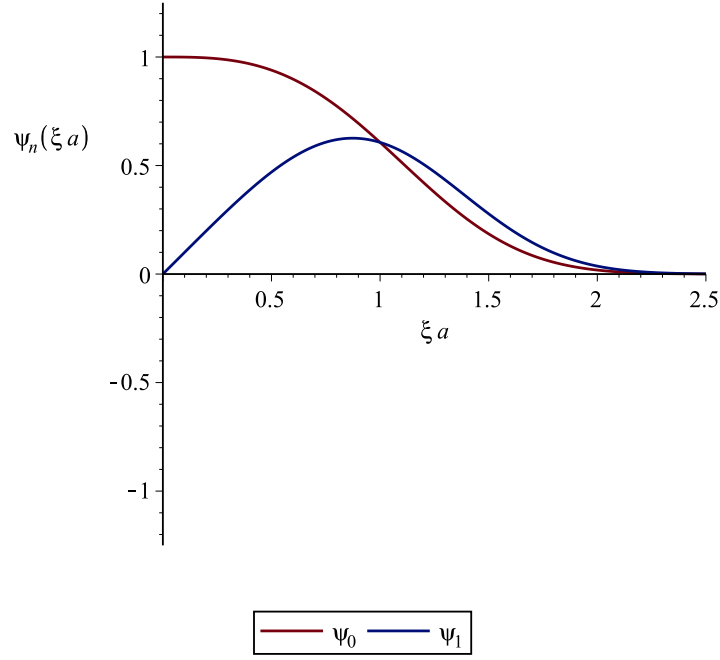}
	\caption{The first three wave functions for $k=0$, $\omega=0$, and $\Lambda = -|\Lambda|$. Note that there is no $\Psi_{2}$ because $\alpha \sim k^{2}$ and we must have $\alpha \in \mathbb{C}^{*}$.}
	\label{fig:WDE_Fig5}
\end{figure}

\begin{figure}%[htbp]
	%\centering
		\includegraphics[scale=0.50]{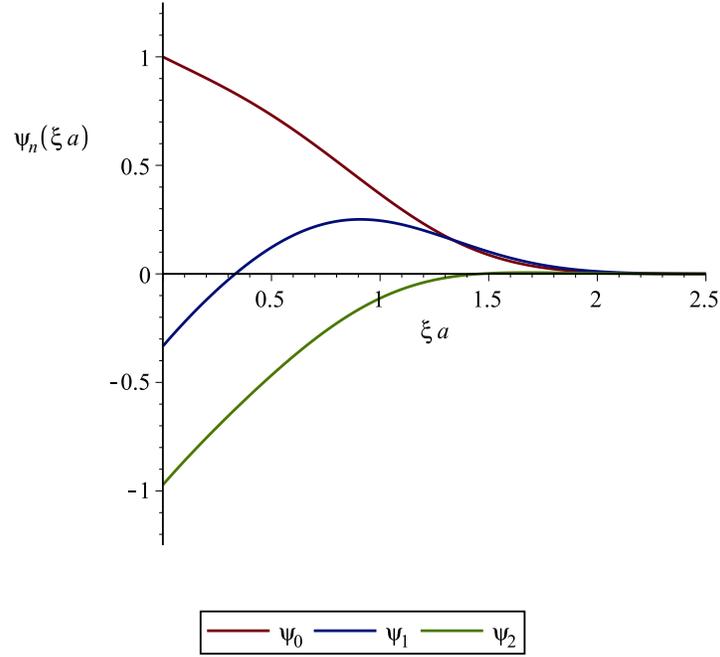}
	\caption{The first three wave functions for $k=1$, $\omega=0$, and $\Lambda = -|\Lambda|$.}
	\label{fig:WDE_Fig6}
\end{figure}

Note that for $\Lambda < 0$ we have bounded states as shown in Fig.~\ref{fig:WDE_Fig3}. From Figs.~\ref{fig:WDE_Fig4}, \ref{fig:WDE_Fig5}, and \ref{fig:WDE_Fig6}, we see that the solutions are all finite in the limit $a \rightarrow \infty$. It is worth calling attention to the fact that the behavior obtained for the wave functions at large values of $a$ guarantees that the correspondence with the classical theory is preserved at this regime.
%
%%%%%%%%%%%%%%%%%%%%%%%%%%%%%%%%%%%%%%%%%%%%%%%%%%%%%%%%%%%%%%%%%%%%%%%%%%%%%%%%%%%%%%%%%%%%%% Solution of the WDW equation for \texorpdfstring{$\omega=1/3$}{w=1/3}
%
\section{Solution of the WDW equation for \texorpdfstring{$\omega=1/3$}{w=1/3}}
\begin{figure}%[htbp]
	%\centering
		\includegraphics[scale=0.50]{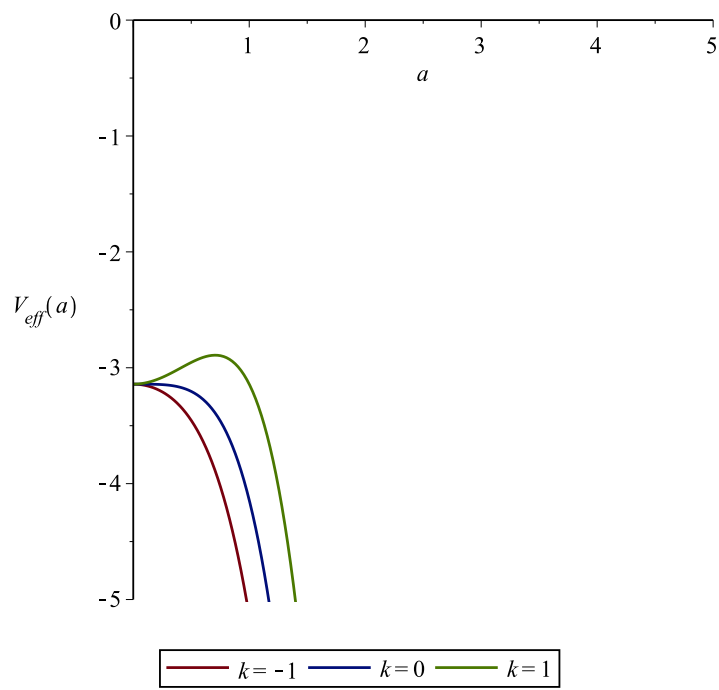}
	\caption{The effective potential energy, $V_{eff}(a)$, for $\omega=1/3$ and $\Lambda > 0$.}
	\label{fig:WDE_Fig7}
\end{figure}

\begin{figure}%[htbp]
	%\centering
		\includegraphics[scale=0.50]{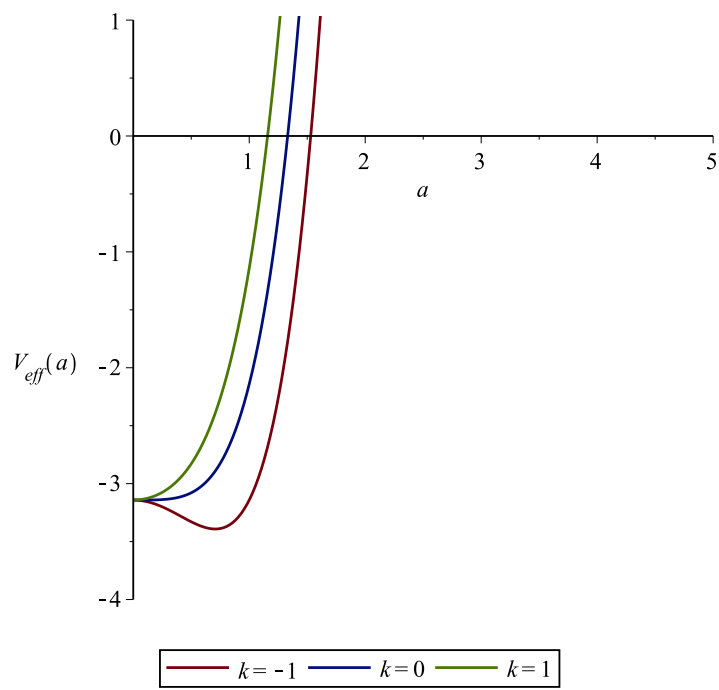}
	\caption{The effective potential energy, $V_{eff}(a)$, for $\omega=1/3$ and $\Lambda = -|\Lambda|$.}
	\label{fig:WDE_Fig8}
\end{figure}

Now, let us assume that $\omega=1/3$, in which case the WDW equation is rewritten as
\begin{equation}
-\hbar^{2}\frac{d^{2}\Psi(a)}{da^{2}}+V_{eff}(a)\Psi(a)=0\ ,
\label{eq:WDE_rad_mov_2}
\end{equation}
where
\begin{equation}
V_{eff}(a)=\frac{9\pi^{2}c^{6}k}{4G^{2}}a^{2}-\frac{6\pi^{3}c^{2}\rho_{r0}a_{0}^{4}}{G}-\frac{3\pi^{2}c^{6}\Lambda}{4G^{2}}a^{4}\ ,
\label{eq:WDE_effective_potential_energy_WDE_w=1/3}
\end{equation}
the behavior of which is shown in Figs.~\ref{fig:WDE_Fig7} and \ref{fig:WDE_Fig8}, for positive and negative values of the cosmological constant, respectively.

It is important to call attention to the fact that for $\omega=1/3$, $\Lambda > 0$, and $k=1$ (closed universe) a tunnelling effect can occur. For $k=0,-1$, thi is no this a possibility. As for $\Lambda < 0$, there is a possibility to create a quantum state by a tunnelling process.

to solve Eq.~(\ref{eq:WDE_rad_mov_2}), let us introduce the dimensionless variable given by Eq.~(\ref{eq:WDE_x}) together with the definitions given by Eqs.~(\ref{eq:WDE_beta_mov_2}) and (\ref{eq:WDE_Omega_mov_2}). Let us also assume that the wave function $\Psi$ is given by Eq.~(\ref{eq:WDE_form_sol_mov_3}), with the parameter $\gamma$ defined as in Eq.~(\ref{eq:WDE_gamma_mov_3}).

Thus, following straightforwardly what was done in the previous, we arrive at an equation for $P(x)$, which is given by
\begin{equation}
\frac{d^{2}P}{dx^{2}}-(\gamma+3x^{2})\frac{dP}{dx}+\left(\frac{6\pi^{3}c^{2}\rho_{r0}a_{0}^{4}}{\xi^{2}\hbar^{2}G}+\frac{\gamma^{2}}{4}-3x\right)P=0\ .
\label{eq:WDE_rad_mov_4}
\end{equation}
Equation (\ref{eq:WDE_rad_mov_4}) is similar to the triconfluent Heun equation shown in Eq.~(\ref{eq:WDE_Triconfluent_Heun_Canonical}). Thus, its general solution can be written as
\begin{eqnarray}
\Psi(a) & = & N\ \mbox{e}^{-\frac{1}{2}(\xi^{3} a^{3}+\gamma \xi a)}\ P(\xi a)\nonumber\\
& = & N\ \mbox{e}^{-\frac{1}{2}(\xi^{3} a^{3}+\gamma \xi a)}\ \mbox{HeunT}(\alpha,\beta,\gamma;\xi a)\ ,
\label{eq:WDE_rad_form_polyn_sol_mov_4}
\end{eqnarray}
where the quantity $N$ is a constant to be determined and the parameters $\alpha$, $\beta$, and $\gamma$ are given by the following expressions:
\begin{equation}
\alpha=\left(\frac{\hbar G}{\pi}\right)^{\frac{2}{3}}\left(\frac{3}{|\Lambda|}\right)^{\frac{1}{3}}\frac{6\pi^{3}\rho_{r0}a_{0}^{4}}{\hbar^{2}G}+\frac{9c^{4}k^{2}}{16}\left(\frac{3\pi}{\hbar G |\Lambda|}\right)^{\frac{4}{3}}\ ,
\label{eq:WDE_rad_alpha_WheelerDeWitt}
\end{equation}
\begin{equation}
\beta=0\ ,
\label{eq:WDE_rad_beta_WheelerDeWitt}
\end{equation}
\begin{equation}
\gamma=\frac{3c^{2}k}{2}\left(\frac{3\pi}{\hbar G |\Lambda|}\right)^{\frac{2}{3}}\ .
\label{eq:WDE_rad_gamma_WheelerDeWitt}
\end{equation}
Note that in the present case $\beta=0$, and thus there is no polynomial solution.
%
%%%%%%%%%%%%%%%%%%%%%%%%%%%%%%%%%%%%%%%%%%%%%%%%%%%%%%%%%%%%%%%%%%%%%%%%%%%%%%%%%%%%%%%%%%%%%% Solution of the WDW equation for \texorpdfstring{$\omega=-1$}{w=-1}
%
\section{Solution of the WDW equation for \texorpdfstring{$\omega=-1$}{w=-1}}
\begin{figure}%[htbp]
	%\centering
		\includegraphics[scale=0.50]{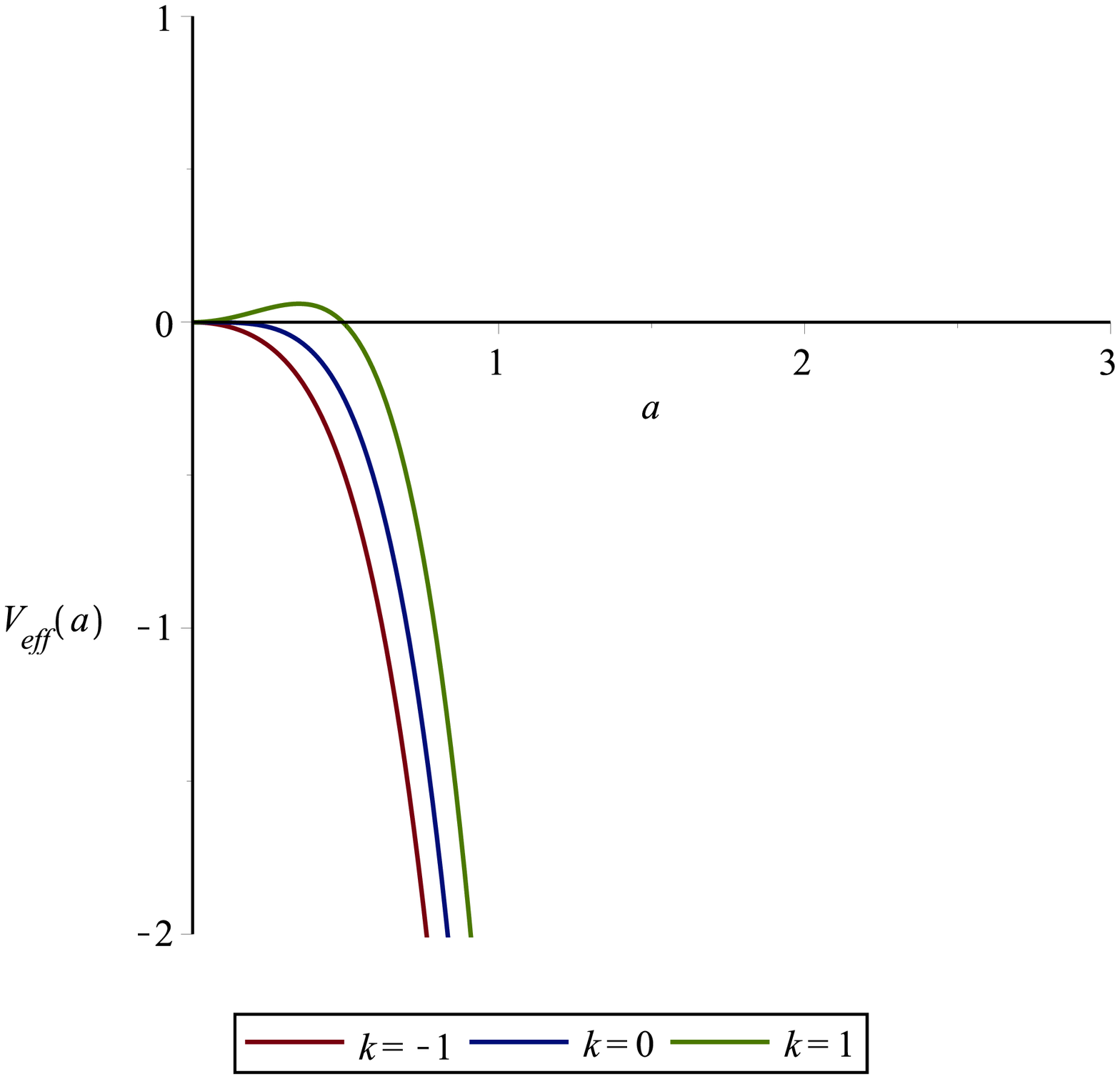}
	\caption{The effective potential energy, $V_{eff}(a)$, for $\omega=-1$ and $\Lambda > 0$.}
	\label{fig:WDE_Fig9}
\end{figure}

\begin{figure}%[htbp]
	%\centering
		\includegraphics[scale=0.50]{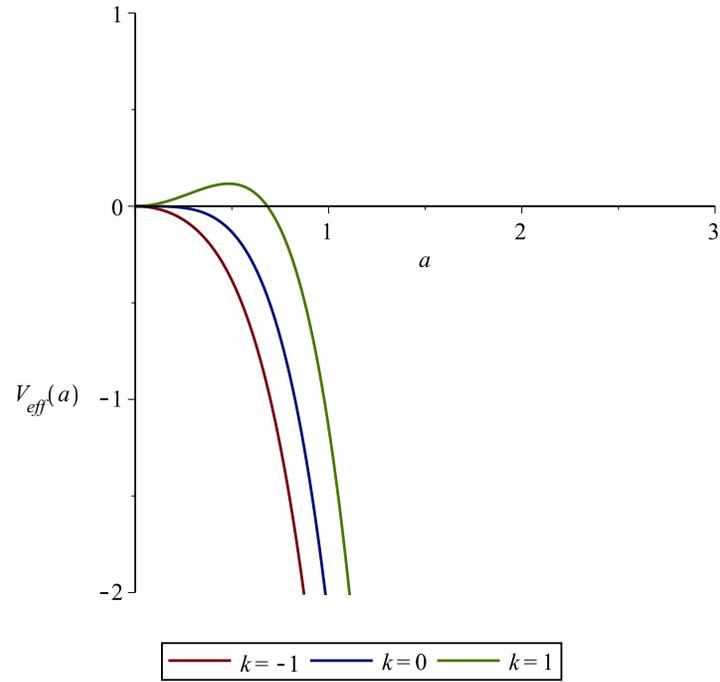}
	\caption{The effective potential energy, $V_{eff}(a)$, for $\omega=-1$ and $\Lambda = -|\Lambda|$.}
	\label{fig:WDE_Fig10}
\end{figure}

For $\omega=-1$, the WDW equation turns into
\begin{equation}
-\hbar^{2}\frac{d^{2}\Psi(a)}{da^{2}}+V_{eff}(a)\Psi(a)=0\ ,
\label{eq:WDE_vac_mov_2}
\end{equation}
where
\begin{equation}
V_{eff}(a)=\frac{9\pi^{2}c^{6}k}{4G^{2}}a^{2}-\left(\frac{6\pi^{3}c^{2}\rho_{v0}}{G}+\frac{3\pi^{2}c^{6}\Lambda}{4G^{2}}\right)a^{4}\ ,
\label{eq:WDE_effective_potential_energy_WDE_w=-1}
\end{equation}
the behaviors of which are shown in Figs.~\ref{fig:WDE_Fig9} and \ref{fig:WDE_Fig10}, for positive and negative values of the cosmological constant, respectively.

In the present case, for $\Lambda > 0$ and $\Lambda < 0$, and taking into account a closed universe, the probability for a tunnel through the barrier to occur is different from zero. For the flat and open universes, there is no possibility of the occurrence of such a phenomenon.

To solve Eq.~(\ref{eq:WDE_vac_mov_2}), once again we will introduce the dimensionless variable as was done in Eq.~(\ref{eq:WDE_x}), where now the coefficient $\xi$ is given by
\begin{equation}
\xi=\left(-\frac{8\pi^{3}c^{2}\rho_{v0}}{3\hbar^{2}G}+\frac{\pi^{2}c^{6}|\Lambda|}{3\hbar^{2}G^{2}}\right)^{\frac{1}{6}}\ .
\label{eq:WDE_vac_tau_mov_2}
\end{equation}
Let us also assume that the wave function, $\Psi$, can be written in accordance with Eq.~(\ref{eq:WDE_form_sol_mov_3}), where now the dimensionless parameter $\gamma$ is given by
\begin{equation}
\gamma=\frac{3\pi^{2}c^{6}k}{2\hbar^{2}G^{2}}\left(-\frac{8\pi^{3}c^{2}\rho_{v0}}{3\hbar^{2}G}+\frac{\pi^{2}c^{6}|\Lambda|}{3\hbar^{2}G^{2}}\right)^{-\frac{2}{3}}\ .
\label{eq:WDE_vac_gamma_mov_3}
\end{equation}
Then, Eq.~(\ref{eq:WDE_vac_mov_2}) turns into
\begin{equation}
\frac{d^{2}P}{dx^{2}}-(\gamma+3x^{2})\frac{dP}{dx}+\left(\frac{\gamma^{2}}{4}-3x\right)P=0\ ,
\label{eq:WDE_vac_mov_4}
\end{equation}
which is similar to the triconfluent Heun equation given by Eq.~(\ref{eq:WDE_Triconfluent_Heun_Canonical}). Thus, as in the previous cases, its general solution can be written as
\begin{eqnarray}
\Psi(a) & = & N\ \mbox{e}^{-\frac{1}{2}(\xi^{3} a^{3}+\gamma \xi a)}\ P(\xi a)\nonumber\\
& = & N\ \mbox{e}^{-\frac{1}{2}(\xi^{3} a^{3}+\gamma \xi a)}\ \mbox{HeunT}(\alpha,\beta,\gamma;\xi a)\ ,
\label{eq:WDE_vac_form_polyn_sol_mov_4}
\end{eqnarray}
where the quantity $N$ is a constant to be determined and the parameters $\alpha$, $\beta$, and $\gamma$ are given by the following expressions:
\begin{equation}
\alpha=\frac{\gamma^{2}}{4}\ ,
\label{eq:WDE_vac_alpha_WheelerDeWitt}
\end{equation}
\begin{equation}
\beta=0\ ,
\label{eq:WDE_vac_beta_WheelerDeWitt}
\end{equation}
\begin{equation}
\gamma=\frac{3\pi^{2}c^{6}k}{2\hbar^{2}G^{2}}\left(-\frac{8\pi^{3}c^{2}\rho_{v0}}{3\hbar^{2}G}+\frac{\pi^{2}c^{6}|\Lambda|}{3\hbar^{2}G^{2}}\right)^{-\frac{2}{3}}\ .
\label{eq:WDE_vac_gamma_WheelerDeWitt}
\end{equation}
As in the radiation-dominated universe, in the present case, $\beta=0$ which means that there is no polynomial solution.
%
%%%%%%%%%%%%%%%%%%%%%%%%%%%%%%%%%%%%%%%%%%%%%%%%%%%%%%%%%%%%%%%%%%%%%%%%%%%%%%%%%%%%%%%%%%%%%% Conclusions
%
\section{Conclusions}
We have shown that in the Friedmann-Robertson-Walker universe, for the spatially closed, flat, and open geometries and taking into account the content of energy corresponding to matter, radiation, and vacuum, all analytical solutions of the Wheeler-DeWitt equation are given in terms of the triconfluent Heun functions. In particular for a universe dominated by matter, the wave function reduces to polynomials in powers of the scalar factor multiplied by an exponential which also depends on the scale factor.

The graphs of the first three wave functions for $k=-1,0,+1$, in a matter dominated universe, are shown in Figs.~\ref{fig:WDE_Fig4}, \ref{fig:WDE_Fig5}, and \ref{fig:WDE_Fig6}, respectively, from which we can see how sensible the solutions are with respect to the geometry. As for the energy spectrum, it corresponds to an infinite sequence of discrete levels, equally spaced, proportional to the square root of the cosmological constant.

In the literature, some investigations concerning the meaning of the solutions of the WDW equation \cite{PhysLettB.748.361}, as well as the transition amplitudes and boundary conditions which should be applied \cite{PhysLettB.117.25}, are based on the solutions of this equation obtained in the classical limit $(a >> 1)$ and in the very early universe $(a << 1)$. With these restrictions some progress has been made in understanding these issues.

In the present paper, we obtained the analytical solutions of the WDW equation valid for all values of the scale factor. Thus, in principle, we can use these solutions to understand the problem concerning the interpretation of the wave function, to find the more appropriate boundary conditions, to investigate the transition amplitudes, and to predict the realization of different stages of the evolution of the universe, but now for all values of the scale factor, and not only very close to the classical limit and very far from this regime.

Therefore, with the solutions of the WDW equation, obtained without the restrictions with respect to the values to be assumed by the scale factor, we believe that, in principle, it would be possible to better understand the issues raised as compared to the scenario in which only asymptotic values of the scale factor, $a << 1$ and $a >> 1$, are taken into account. In this way, we expect that some more progress in understanding the role played by the WDW equation in quantum cosmology could be made.
%
%%%%%%%%%%%%%%%%%%%%%%%%%%%%%%%%%%%%%%%%%%%%%%%%%%%%%%%%%%%%%%%%%%%%%%%%%%%%%%%%%%%%%%%%%%%%%% acknowledgments
%
\begin{acknowledgments}
The authors would like to thank Conselho Nacional de Desenvolvimento Cient\'{i}fico e Tecnol\'{o}gico for partial financial support. H. S. V. is funded through the research Project No. 140612/2014-9. V. B. B. is partially supported through the research Project No. 304553/2010-7.
\end{acknowledgments}
%
%%%%%%%%%%%%%%%%%%%%%%%%%%%%%%%%%%%%%%%%%%%%%%%%%%%%%%%%%%%%%%%%%%%%%%%%%%%%%%%%%%%%%%%%%%%%%% thebibliography
%

%
%%%%%%%%%%%%%%%%%%%%%%%%%%%%%%%%%%%%%%%%%%%%%%%%%%%%%%%%%%%%%%%%%%%%%%%%%%%%%%%%%%%%%%%%%%%%%%
%
\end{document}